\documentclass[prl,aps,twocolumn,superscriptaddress]{revtex4}

\usepackage{epsfig}
\usepackage{natbib}
\usepackage{mathrsfs}
\usepackage{amsmath}

\begin{document}

\title{Gapless Spin Excitations in the Field-Induced Quantum Spin Liquid
Phase of $\alpha$-RuCl$_3$}

\author{Jiacheng Zheng}
\affiliation{Department of Physics and Beijing Key Laboratory of
Opto-electronic Functional Materials $\&$ Micro-nano Devices, Renmin
University of China, Beijing, 100872, China}
\affiliation{Department of Physics, Beijing Jiaotong University, Beijing
100044, China}

\author{Kejing Ran}
\affiliation{National Laboratory of Solid State Microstructures
and Department of Physics, Nanjing University, Nanjing 210093, China}

\author{Tianrun Li}
\affiliation{Department of Physics and Beijing Key Laboratory of
Opto-electronic Functional Materials $\&$ Micro-nano Devices, Renmin
University of China, Beijing, 100872, China}

\author{Jinghui Wang}
\affiliation{National Laboratory of Solid State Microstructures
and Department of Physics, Nanjing University, Nanjing 210093, China}

\author{Pengshuai Wang}
\affiliation{Department of Physics and Beijing Key Laboratory of
Opto-electronic Functional Materials $\&$ Micro-nano Devices, Renmin
University of China, Beijing, 100872, China}

\author{Bin Liu}
\affiliation{Department of Physics, Beijing Jiaotong University, Beijing
100044, China}

\author{Zhengxin Liu}
\affiliation{Department of Physics and Beijing Key Laboratory of
Opto-electronic Functional Materials $\&$ Micro-nano Devices, Renmin
University of China, Beijing, 100872, China}

\author{B. Normand}
\affiliation{Laboratory for Neutron Scattering and Imaging, Paul Scherrer
Institute, \\ CH-5232 Villigen-PSI, Switzerland}

\author{Jinsheng Wen}
\email{jwen@nju.edu.cn}
\affiliation{National Laboratory of Solid State Microstructures and Department
of Physics, Nanjing University, Nanjing 210093, China}
\affiliation{Innovative Center for Advanced Microstructures, Nanjing University,
Nanjing 210093, China}

\author{Weiqiang Yu}
\email{wqyu\_phy@ruc.edu.cn}
\affiliation{Department of Physics and Beijing Key Laboratory of
Opto-electronic Functional Materials $\&$ Micro-nano Devices, Renmin
University of China, Beijing, 100872, China}
\affiliation{Department of Physics and Astronomy, Shanghai Jiaotong
University, Shanghai 200240, China and \\ Collaborative Innovation Center
of Advanced Microstructures, Nanjing 210093, China}



\begin{abstract}

$\alpha$-RuCl$_3$ is a leading candidate material for the
observation of physics related to the Kitaev quantum spin liquid (QSL).
By combined susceptibility, specific-heat, and nuclear-magnetic-resonance
measurements, we demonstrate that $\alpha$-RuCl$_3$ undergoes a quantum phase
transition to a QSL in a magnetic field of 7.5 T applied in the $ab$ plane.
We show further that this high-field QSL phase has gapless spin excitations
over a field range up to 16 T. This highly unconventional result, unknown
in either Heisenberg or Kitaev magnets, offers insight essential to
establishing the physics of $\alpha$-RuCl$_3$.

\end{abstract}

\maketitle

The quantum spin liquid (QSL) is an exotic state of matter with long-range
coherence but with no spontaneous breaking of translational or spin-rotational
symmetry down to zero temperature \cite{Balents_nature464_199}. Such a
state in two or higher dimensions has implications for phenomena ranging
from high-temperature superconductivity \cite{Anderson1987} to quantum
computation \cite{Kitaev_ap_2003, Nayak_RMP.80.1083}. QSLs have long
been sought in systems with strong geometric frustration
\cite{Anderson1973153,Kanoda_PRL_2003,Han_nature492_406,Greedan_RMP.82.53},
where magnetic order is destroyed by quantum fluctuations in a highly
degenerate ground manifold. A more recent avenue to QSL formation is by
competing interactions with combined spin and spatial anisotropies, as in
the Kitaev model, where both gapped and gapless QSLs are realized exactly
in a honeycomb-lattice spin-1/2 system \cite{Kitaev_ap_2006}.

\begin{figure}[t]
\includegraphics[width=8cm]{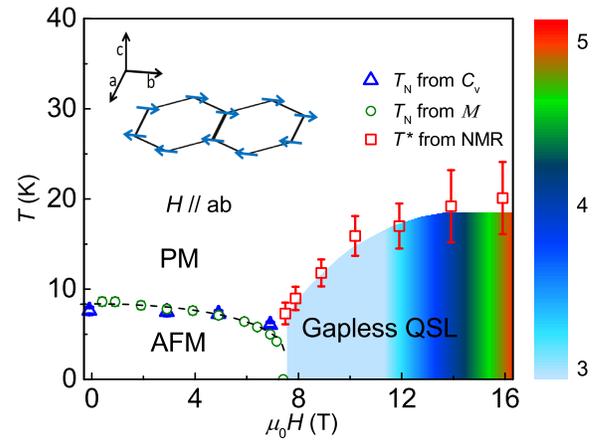}
\caption{\label{pd1} {\bf Magnetic phase diagram of $\alpha$-RuCl$_3$ with
field applied in the $ab$ plane.} $T_N$ is determined from magnetization and
specific-heat data (Fig.~2). In the QSL phase, the color map represents the
exponent, $\alpha$, determined from the power-law form of the NMR spin-lattice
relaxation rate, $1/^{35}T_1 \propto T^\alpha$ (Fig.~4). $T^*$ represents the
upper limit of the gapless low-$T$ regime. Inset: schematic representation
of zero-field zig-zag order in the hexagonal ($ab$) plane. }
\end{figure}

A pure Kitaev Hamiltonian is hard to achieve in real materials. However,
the compounds A$_2$IrO$_3$ (A = Na, Li) \cite{Khaliullin_prl102_017205,
Khaliullin_prl105_027204,Singh_PhysRevLett.108.127203,
Gretarsson_PhysRevLett.110.076402,Chun_np_2015,Kee_conmatphys_2016} and
$\alpha$-RuCl$_3$ \cite{Pollini_PhysRevB.53.12769,Plumb_PhysRevB.90.041112,
Kee_PhysRevB.91.241110,Burch_Prl_2015,Kindo_PhysRevB.91.094422,
Sandilands_PhysRevB.93.075144,Koirzsch_PRL.117.126403} are candidate systems
for significant Kitaev-type interactions. In each case, the $4d$ (Ru$^{3+}$)
or $5d$ (Ir$^{4+}$) ions form a Mott insulator on a honeycomb lattice, whose
localized electrons have an effective spin $j_{\rm eff} = 1/2$ due to strong
spin-orbit coupling \cite{Khaliullin_prl102_017205,Kim_science_2009,
Foyevtsova_PhysRevB.88.035107,Khaliullin_PhysRevB.94.064435,
Haule_PhysRevLett.114.096403}. In $\alpha$-RuCl$_3$ at zero field, a
finite-energy continuum of magnetic excitations \cite{Banerjee_nm15_733}
is suggestive of fractionalized (spinon or Majorana-fermion) excitations
\cite{Coldea_PhysRevB.68.134424,Kitaev_ap_2006,Shankar_PhysRevLett.98.247201,
Moessner_PhysRevLett.112.207203,Moessner_PhysRevB.92.115127,Nasu_np12_912}.
However, the ground states in all cases have ``zig-zag'' magnetic order
\cite{Ye_PhysRevB.85.180403,Choi_PhysRevLett.108.127204,Fletcher_JCSA_1967,
Sears_PhysRevB.91.144420,Johnson_PhysRevB.92.235119,Cao_PRB.93.134423},
indicating the presence of significant non-Kitaev terms, whose exact nature
continues to occupy many authors \cite{Kimchi_PhysRevB.84.180407,
Mazin_PhysRevLett.109.197201,Khaliullin_prl110_097204,Rau_PRL.112.077204,
Yamaji_PRL_2014,Perkins_PRB.90.155126,Winter_PRB_2016,Kim_PRB.93.155143,
Perkins_PRB.94.085109,Wang_arxiv_2016}. While the large $T_N$ in Na$_2$IrO$_3$
\cite{Ye_PhysRevB.85.180403,Choi_PhysRevLett.108.127204} suggests subdominant
Kitaev terms, the relatively low $T_N$ of $\alpha$-RuCl$_3$ has sparked an
intensive search for experimental \cite{Nagler_2016_arxiv,LeeM_PRL.118.187203,
Baek_arxiv_170201671} and theoretical \cite{Janssen_arxiv_2016,Yadav_arxiv_2016}
evidence for ``proximate Kitaev'' behavior.

\begin{figure*}[t]
\includegraphics[width=16.5cm, height=6.8cm]{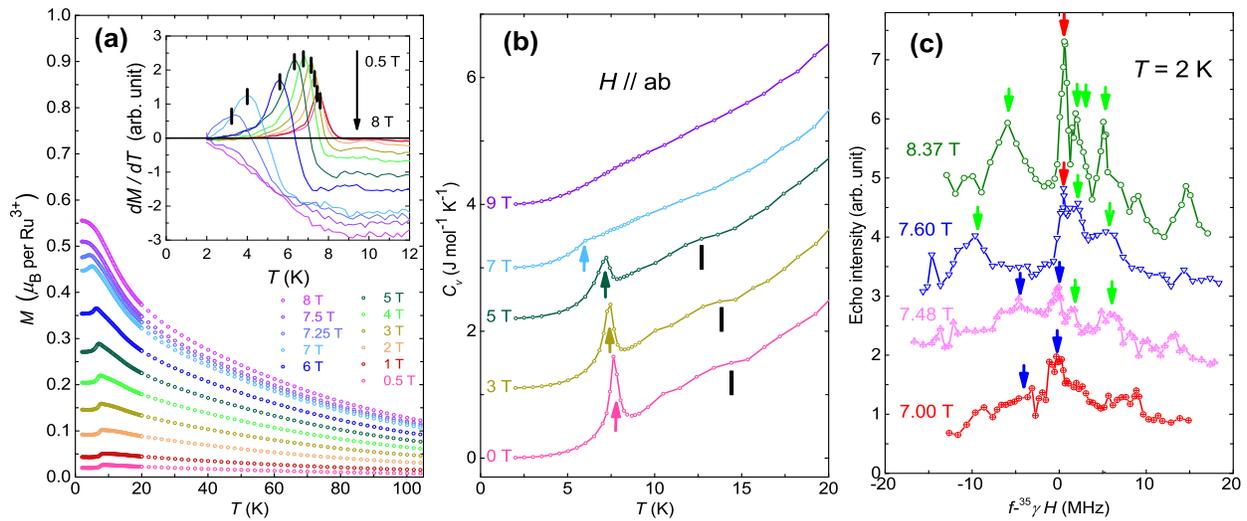}
\caption{\label{suscv2} {\bf Magnetic transition with field
applied in the $ab$ plane.} (a) Magnetization, $M(T)$. Inset: $dM/dT$;
vertical lines mark the peaks, which show $T_N$ for fields up to 7.25 T. (b)
Low-$T$ specific heat, $C_v$. Data are offset for clarity. The transition for
regions with $ABC$ ($AB$) layer stacking is marked by the arrows (vertical
lines). (c) $^{35}$Cl NMR spectra at fields close to 7.5 T, shown at $T = 2$
K. Blue arrows mark peaks characteristic of the low-field, ordered phase,
green arrows the peaks of the high-field, disordered phase, and red arrows
the sharp central peak at which $^{35}K_n$ and $1/^{35}T_1$ were measured.}
\end{figure*}

Here we report a nuclear-magnetic-resonance (NMR) investigation of
high-quality single crystals of $\alpha$-RuCl$_3$. With additional magnetic
susceptibility and specific-heat measurements, we establish the phase diagram
of Fig.~\ref{pd1}. We demonstrate the presence of a field-induced QSL beyond
the quantum phase transition at $\mu_0 H_c \simeq$ 7.5 T. In the field range
between 7.5 T and 16 T, this partially polarized QSL has a spin-lattice
relaxation rate with power-law temperature dependence, indicating effectively
gapless spin excitations with line-node dispersion.

Single crystals of $\alpha$-RuCl$_3$ were grown by chemical vapor
transport. The high quality of this batch of crystals is demonstrated in
Ref.~\cite{WenJS_prl_2016} and x-ray characterization of the NMR sample,
shown in Sec.~S1 of the Supplemental Material (SM) \cite{sm}, demonstrated
that it is a single domain and free of twinning. Magnetization measurements
were performed in a 9T SQUID and the specific heat measured in a Quantum
Design PPMS. $^{35}$Cl NMR spectra were collected by the spin-echo technique
and the spin-lattice relaxation rate measured by the inversion-recovery
method, as shown in Sec.~S2 of the SM \cite{sm}. The spin-recovery exponent,
$\beta = 1$ in the paramagnetic state ($T > 20$ K in Fig.~1), also indicates
a very high sample quality.

The magnetization, $M(T)$, is shown for a range of applied fields in
Fig.~\ref{suscv2}(a), with its primary features, emphasized by the low-$T$
derivative, $dM/dT$, shown in the inset. At zero field, $\alpha$-RuCl$_3$ has
zig-zag antiferromagnetic (AFM) order (inset, Fig.~\ref{pd1}) below $T_N
\simeq$ 7.5 K (14 K) for crystals with ABC (AB) stacking along the $c$-axis
\cite{Cao_PRB.93.134423}. The sharp phase transition at 7.5 K in both $dM/dT$
and the specific heat, $C_v$ [Fig.~\ref{suscv2}(b)], demonstrates the very
high crystal quality. Our samples are almost exclusively ABC-stacked, with
only a small admixture of AB stacking discernible through the weak anomaly
in $C_v$. Our $C_v$ data include the phonon contribution; due to concerns
over its subtraction and over the suitability of specific heat for this
purpose, we do not attempt to use our data to analyze the magnetic response.

\begin{figure*}[t]
\includegraphics[width=14cm, height=6cm]{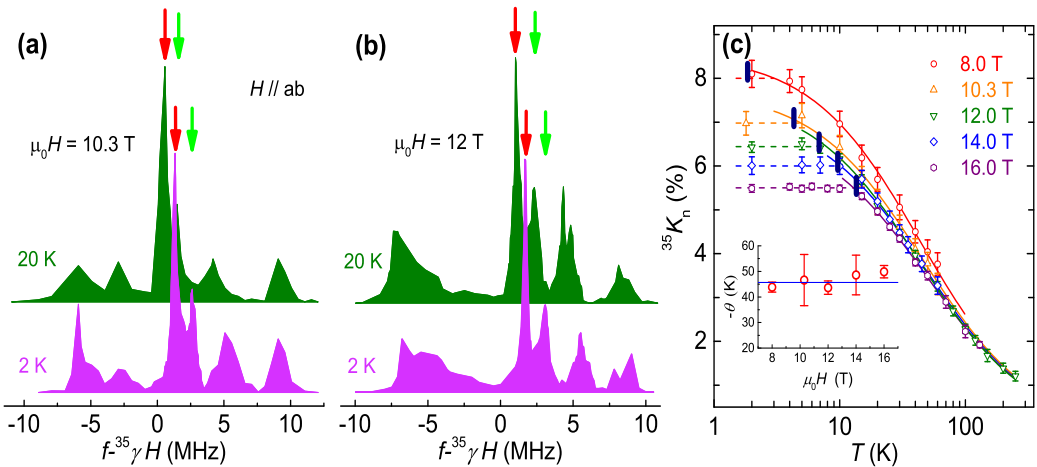}
\caption{\label{spec3}{\bf $^{35}$Cl NMR spectra and Knight shift for in-plane
fields.} $^{35}$Cl spectra measured at (a) 10.3 T and (b) 12 T, shown for
$T = 2$ K and 20 K. $\Delta f = f - ^{35}\gamma H$ is the frequency shift,
where $f$ is the measured frequency and $^{35}\gamma$ = 4.171 MHz/T is the
gyromagnetic ratio of $^{35}$Cl. The splitting of the sharp central peak (red
and green arrows) is a consequence of the three inequivalent Cl$^{-}$ sites,
two of which are not separated at this field angle. The separation of the
broader satellite peaks from the center peaks changes little with field. The
peak widths are determined by quadrupolar effects \cite{Abragam}, discussed
in Sec.~S5 of the SM \cite{sm}. (c) NMR Knight shift, $^{35}K_n(T)$, measured
at the central peak and shown for different field values. $^{35}K_n$ is
calculated from $\Delta f$ after subtracting all $T$-independent quadrupolar
and orbital contributions (Sec.~S5 \cite{sm}). Solid lines are Curie-Weiss
fits to the high-$T$ data, of the form $^{35}K_n = a/(T - \theta)$; we find
that $\theta = -45 \pm 10$ K is almost field-independent (inset). Dotted
lines indicate a ``level-off'' behavior of $^{35}K_n$ at low temperatures
(marked by the vertical lines).}
\end{figure*}

Fields applied in the $ab$-plane suppress $T_N$. This effect is especially
strong from 7 T to 7.25 T, leading up to a field-induced quantum phase
transition (QPT) at $\mu_0 H_c \simeq$ 7.5 T. The QPT is observed clearly
both in the disappearance of the peaks in $dM/dT$ and $C_v$ and in the
dramatic changes in the NMR spectra [Fig.~\ref{suscv2}(c)], which we discuss
next. We find no anisotropy in $M$ and $C_v$ as the field is rotated in the
$ab$-plane. The NMR measurements shown in Fig.~\ref{suscv2}(c), and also
Fig.~\ref{spec3}, are fully representative of the generic in-plane response,
as we demonstrate in Secs.~S3 and S4 of the SM \cite{sm}.

We collect the evidence that the phase at $H > H_c$ is a QSL. First, the
vanishing peaks in both $dM/dT$ and $C_v$ (Fig.~\ref{suscv2}) demonstrate
the absence of magnetic order. Second, the $^{35}$Cl line shapes for fields
at and above 7.6 T contain none of the peaks corresponding to an ordered Cl
environment. The $^{35}$Cl NMR spectra below 7.6 T [Fig.~\ref{suscv2}(c)] are
very broad at low $T$, consistent in the absence of domain and twinning
effects only with AFM order. By contrast, several sharp peaks are clearly
resolvable for fields above 7.6 T in Figs.~\ref{suscv2}(c), \ref{spec3}(a),
and \ref{spec3}(b). The center peak at $\Delta f \approx$ 0 ($\gamma H
\approx$ 43.5 MHz) has a FWHM height of 0.5 MHz at 10.3 T at both $T = 20$ K
and 2 K [Fig.~\ref{spec3}(a)], and shows no significant changes at 12 T
[Fig.~\ref{spec3}(b)]. Such narrow linewidths, and in particular their
invariance upon cooling below 20 K, indicate a complete absence of magnetic
order.

Further evidence is provided by the Knight shift, $^{35}K_n$, of the center
peak, shown in Fig.~\ref{spec3}(c). From its large values (6-7$\%$) below
10 K, the hyperfine field is strong at the $^{35}$Cl site and magnetic order,
if present, is very unlikely to be missed by NMR. This statement remains
true even for incommensurate or large-unit-cell ordered phases that are
possible in a field \cite{Janssen_arxiv_2016}. Further, a Curie-Weiss (CW)
fit to the high-temperature part of the data holds down to 2 K at fields
of 8 T, with the CW temperature, $\theta \approx - 45$ K [inset,
Fig.~\ref{spec3}(c)], supporting the absence of magnetic order. Because
the NMR spectra show no evidence that the magnetic QPT is accompanied by a
structural transition, which would cause much more dramatic peak shifts in
Fig.~\ref{suscv2}(c), we conclude that the field-induced disordered phase
is indeed a QSL.

Unlike $T_N$, the value of $H_c$ required to suppress magnetic order in a
crystal with AB stacking is also around 8 T \cite{Johnson_PhysRevB.92.235119}.
The similarity of $H_c$ values for ABC and AB stacking suggests that the
field-induced suppression of magnetic order, and by extension the properties
of the high-field QSL phase, are primarily two-dimensional (2D) in nature,
rather than depending on interlayer coupling.

\begin{figure}[t]
\includegraphics[width=7.5cm]{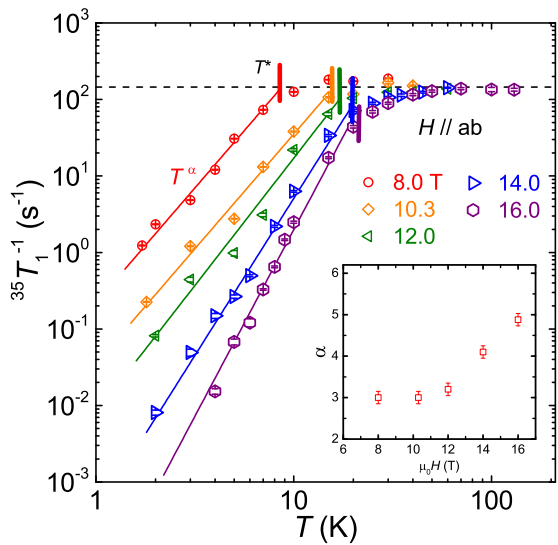}
\caption{\label{slrr4}{\bf NMR spin-lattice relaxation rates.} $1/^{35}T_1$
shown for five different applied in-plane fields. At $T > 40$ K, $1/^{35}T_1$
approaches a constant value, $1/^{35}T_1 \simeq$ 145 s$^{-1}$, for all fields
(dashed line). At low $T$, $1/^{35}T_1$ follows a power-law $T$-dependence;
solid lines are fits to the form $1/^{35}T_1 = b T^\alpha$ over one decade in
$T$. The exponent $\alpha$ (inset) and temperature scale $T^*$ (vertical lines)
are also shown in Fig.~\ref{pd1}.}
\end{figure}

The spin excitations of the high-field QSL state are probed by the
spin-lattice relaxation rate. We measure $1/^{35}T_1$ for each field at
the sharp central peak marked by the red arrows in Figs.~\ref{suscv2} and
\ref{spec3}, and show the results in Fig.~\ref{slrr4}. Above 40 K, $1/^{35}T_1$
remains constant in both $T$ and $H$, which is typical of decorrelated spins
at temperatures beyond their interaction energy scale. At 8 T, $1/^{35}T_1$
drops sharply on cooling below 10 K, and follows a power-law $T$-dependence,
$1/^{35}T_1 = b T^\alpha$, to the bottom of our measurement range ($T = 1.5$ K).
The same behavior holds at all fields up to our maximum of 16 T. The exponent
$\alpha$, shown in the inset of Fig.~\ref{slrr4}, remains constant at $\alpha
 = 3$ for fields between 8 and 12 T, and then increases to $\alpha = 5$ at 16
T. While exponents above 5 suggest the opening of a full gap, the reliable
extraction of power-law behavior with exponents as small as $\alpha = 3$
demonstrates that the QSL in the field range $8 < \mu_0 H < 16$ T has either
gapless spin excitations or an anomalously small gap far below 2 K.

We extract also the characteristic temperature, $T^*$, marking the upper
limit of the low-$T$, power-law regime, which is shown in Fig.~\ref{slrr4}
and also in Fig.~\ref{pd1}. The increase of $T^*$ with field demonstrates
that the phase of coherent QSL dynamics becomes increasingly robust, at
least to 16 T. This occurs in tandem with an increasing field-driven
polarization, observed in both $M$ and $^{35}K_n$, which leaves a decreasing
component of the spin available to participate in the QSL. The related
average moment per Ru ion has been measured directly
\cite{Johnson_PhysRevB.92.235119} as 0.56$\mu_B$ at 8 T, 0.87$\mu_B$ at
16 T, and 1.22$\mu_B$ at 60 T. For comparison, the static moment in the
zig-zag AFM state at $H = 0$ is 0.4$\mu_B$. We note that the recovery
exponent, $\beta$, decreases at $T < T^*$, suggesting that the QSL dynamics
are very sensitive to any weak disorder. However, as discussed in Sec.~S2
of the SM \cite{sm}, our results contain no evidence that the QSL state
itself could be a consequence of disorder, and verify rather its intrinsic
nature.

Two recent experiments have also investigated $\alpha$-RuCl$_3$
in fields exceeding $H_c$. Thermal conductivity measurements show a
prominent low-$T$ peak whose magnitude grows linearly with $H - H_c$
\cite{LeeM_PRL.118.187203}, indicating a gapless and linearly dispersive
excitation. By contrast, a different NMR study reports gapped behavior
\cite{Baek_arxiv_170201671}, although the field is applied at an angle
30$^\circ$ out of the $ab$-plane. There is increasing evidence that
different field orientations lead to qualitatively different properties
in $\alpha$-RuCl$_3$, as expected from the strong spin anisotropy (our own
NMR investigation is presented in Sec.~S4 of the SM \cite{sm}). However, the
gaps reported by these authors are deduced only over a limited temperature
range that, crucially, does not extend to $T = 0$, and the finite $1/T_1$
observed at low $T$ is not consistent with a spin gap.

To interpret our results, we consider the definition $1/T_1 T = {\rm lim}
_{\omega \to 0} \sum_{q} A_{\rm hf}(q) \, {\rm Im} \chi(q,\omega)/\omega$, where
$\chi(q, \omega)$ is the dynamical susceptibility and $A_{\rm hf}(q)$ the
hyperfine coupling. For a conventional magnetic system, this may be
reexpressed as $1/T_1 \sim \int_{0}^{\infty} \rho^2(E) n(E) [1 + n(E)] dE$,
where $\rho(E)$ is the magnon density of states (d.o.s.) and $n(E)$ the Bose
distribution function, and this makes $1/T_1$ a sensitive probe of low-energy
spin dynamics. Our result $1/T_1 \sim T^3$ suggests that the density of
magnetic states follows $\rho(E) \propto E$. In condensed matter, such a
linear d.o.s. is more familiar from fermionic excitations, including line
nodes in the gap at the Fermi surface of a 3D superconductor~\cite{Monien}
and the 2D Dirac-cone dispersion~\cite{Dora}.

In conventional (Heisenberg) quantum magnets, the partial polarization is
distributed uniformly and becomes complete at a finite saturation field,
$H_s$. All spin excitations are bosonic and for $H > H_s$ they have a gap
$\Delta \propto H - H_s$. Spatially anisotropic and frustrated systems show
magnetization plateaus at low or intermediate fields, whose ground states
are in general gapped \cite{Tamaka_PRL_102_2009} with bosonic excitations.
Thus our discovery of gapless spin excitations at intermediate fields in
$\alpha$-RuCl$_3$ represents an extremely unconventional situation.

When a Hamiltonian anisotropic in spin space does not commute with the field,
$H_s \rightarrow \infty$. Although the spatially isotropic Kitaev model has
one gapless and linearly dispersive Majorana fermion at zero field, its spin
excitations, consisting of Majorana fermions coupled to two massive, static
flux quanta \cite{Kitaev_ap_2006,Shankar_PhysRevLett.98.247201}, are gapped.
In Ref.~\cite{Song_prl_2016} it was shown that the spin gap can vanish in a
``generic Kitaev'' system, meaning one in which the symmetry-allowed
Heisenberg ($J$) and off-diagonal ($\Gamma$) terms are present at a
perturbative level, and these authors obtained the highly suggestive result
$1/T_1 \propto T^3$. However, in an applied field, all the modes in both the
pure ($K$) and generic (perturbative $K$-$J$-$\Gamma$) Kitaev models become
gapped. Although a gapless Majorana quasiparticle could be responsible for
the measured thermal conductivity \cite{LeeM_PRL.118.187203}, our NMR results
show unequivocally that the spin excitations themselves are gapless, and thus
such proximate Kitaev physics appears to be excluded.

We review scenarios allowing gapless spin excitations over a range of field
strengths. As a result of its weak interplane interactions, $\alpha$-RuCl$_3$
is effectively a 2D magnet \cite{Banerjee_nm15_733} and therefore its line
nodes are interplane, connecting point nodes in the honeycomb layers. Bosonic
excitations in (and beyond) 2D are either gapped or condense, leading to
magnetic order, whereas gapless and disordered (i.e.~critical) behavior over
a finite field range is unknown. Turning to fermionic excitations, these imply
a fractionalization and deconfinement of spinonic quasiparticles taking place
at the QPT. In the Heisenberg chain and ladder (1D), gapless spinons appear
at incommensurate wave vectors over a range of fields. In 2D, some
spin-orbit-coupled magnetic systems may be represented by fermionic spinons,
whose dispersion has Dirac cones. These cones provide the linear d.o.s. and
remain stable, i.e.~pinned at the Fermi level, over a finite range of fields
applied in specific symmetry directions, whereas fields in generic directions
open a gap (an example is shown in Sec.~S6 of the SM \cite{sm}). For Majorana
fermions, at least two flavors of Majorana cone are required for the system
to host gapless spin excitations.

Concerning a microscopic model exhibiting such exotic physics, the
effective magnetic Hamiltonian of the spin-orbit-coupled Mott insulator has
focused attention on $K$-$J$-$\Gamma$-type models \cite{Rau_PRL.112.077204,
Yamaji_PRL_2014}. Opinions on the terms and parameters describing
$\alpha$-RuCl$_3$ remain strongly divergent. Early efforts using a $J$-$K$
model with ferromagnetic $J$ and AFM $K$ \cite{Banerjee_nm15_733} have
been supplanted by an exchange of signs and longer-ranged $J$ terms
\cite{Winter_PRB_2016,Yadav_arxiv_2016}. Very recent studies
\cite{Nagler_2016_arxiv,Wang_arxiv_2016,WenJS_prl_2016} have turned to
$K$-$\Gamma$ models with parameters far outside the perturbative regime
of Ref.~\cite{Song_prl_2016}. Although the effects of a magnetic field have
yet to be investigated in detail, our own $K$-$\Gamma$ analysis (\cite{rln},
Sec.~S6 of the SM \cite{sm}) indicates robust QSL states whose gapless
fermionic spinons have four dispersion cones for certain field directions
and a very small gap for all in-plane fields.

In summary, we have observed a QPT to a field-induced QSL above $\mu_0 H_c
 = 7.5$ T in $\alpha$-RuCl$_3$. We have shown by NMR measurements that this
state has effectively gapless spin excitations over a broad field range. This
result cannot be reconciled with the behavior of conventional quantum magnets
or of the pure or generic Kitaev QSL. Thus our data suggest fractionalized
spinon excitations in $\alpha$-RuCl$_3$ and set a significant challenge to
theory.

This work was supported by the National Science Foundation of China (Grant
Nos.~11374364, 11374143, and 11674157), by the Ministry of Science and
Technology of China (Grant No.~2016YFA0300504), and by the Fundamental
Research Funds for the Central Universities and the Research Funds of Renmin
University of China (Grant No.~14XNLF08). JCZ and KJR contributed equally to
this study.

\end{document}